\documentclass[prl,showpacs,twocolumn,amsmath,amssymb,floatfix,superscriptaddress]{revtex4-1}

\usepackage{graphicx}
\usepackage{pstricks,pst-node,pst-text,pst-3d}
\usepackage{color}

\newcommand{\pdagger}{{\phantom{\dagger}}}
\newcommand{\dt}{\Delta\tau}
\newcommand{\reff}[1]{Fig.\ \ref{fig:#1}}

\newcommand{\myparagraph}[1]{{\it #1} -- }

\newcommand{\neel}{N\'{e}el}
\newcommand{\TN}{T_{\text{N}}}
\newcommand{\TND}{T_{\text{N}}^{\text{DMFT}}}

\newcommand{\NNcorr}{\langle\hat{\boldsymbol{\sigma}}_i\cdot\hat{\boldsymbol{\sigma}}_j\rangle}

\begin{document}

	\title{Double occupancy as a universal probe for antiferromagnetic correlations\\
	       and entropy in cold fermions on optical lattices}

\author{E.~V.~Gorelik}
\affiliation{Institute of Physics, Johannes Gutenberg University, Mainz, Germany}
\author{T.~Paiva}
\affiliation{Instituto de Fisica, Universidade Federal do Rio de Janeiro, Brazil}
\author{R.~Scalettar}
\affiliation{Department of Physics, UC Davis, USA}
\author{A.~Kl\"umper}
\affiliation{University of Wuppertal, Wuppertal, Germany}
\author{N.~Bl\"umer}
\affiliation{Institute of Physics, Johannes Gutenberg University, Mainz, Germany}

\date{\today}

  \begin{abstract}
  	We verify signatures of antiferromagnetic (AF) correlations in the double occupancy $D$ [Gorelik et al., PRL {\bf 105}, 065301 (2010)] and study their dimensional dependence using direct quantum Monte Carlo in dimensions $d=2,3$ and Bethe Ansatz in $d=1$. We find quantitative agreement with dynamical mean-field theory (DMFT) in the cubic case and qualitative agreement down to $d=1$. As a function of entropy $s=S/(N k_{\text{B}})$, $D$ is nearly universal with respect to $d$; the minimum in $D(s)$ approaches $s\approx \log(2)$ at strong coupling, as predicted by DMFT. Long-range order appears hardly relevant for the current search of AF signatures in cold fermions. Thus, experimentalists need not achieve $s<\log(2)/2$ and should consider lower dimensions, for which the AF effects are larger.
  \end{abstract}
  \maketitle

%%%%%%%%%%%%%%%%%%%%%%%%%%%%%%%%%%%%%%%%%%%%%%%%%%%%%%%%%%%%%%%%%%%%%%%%
% Introduction
%%%%%%%%%%%%%%%%%%%%%%%%%%%%%%%%%%%%%%%%%%%%%%%%%%%%%%%%%%%%%%%%%%%%%%%%

A thorough understanding of materials with strong electronic correlations is not only desirable on intellectual grounds, but also due to their increasing technological importance, e.g., in magnetoresistive and superconducting devices \cite{Tokura03,Dagotto05}. Theoretical investigations of corresponding Hubbard type models using direct analytical techniques, numerical approaches for finite clusters, and the dynamical mean-field theory (DMFT) have already shed light on many strong-coupling phenomena including metal-insulator transitions, heavy-fermion and non-Fermi-liquid behavior, and various types of magnetic and orbital order \cite{AnisimovBook2010}. However, there are still important open questions, most notably regarding high-temperature superconductivity for which so far no mechanism could conclusively be established. In this situation, the recent advent of a novel class of correlated Fermi systems, namely ultracold fermionic atoms (such as $^{40}$K and $^6$Li) on optical lattices, have opened a new promising direction of research: cold atoms are predicted to serve as {\it quantum simulators} for the Hubbard type solid-state Hamiltonians of interest \cite{Zoller05,Esslinger10}.

Indeed, within a few years after the first achievement of quantum degeneracy in (single flavor) fermionic atoms on optical lattices \cite{Koehl05}, the Mott metal-insulator transition (MIT) was observed in two-flavor mixtures, based on signatures in the compressibility \cite{Bloch08_ferm} and a suppression of the integrated double occupancy \cite{Esslinger08_Nature}. As a result, it is now established that the single-band Hubbard model 
\[
  \hat{H} =\! -t \sum_{\langle ij\rangle ,\sigma}  \hat{c}^{\dag}_{i\sigma}
\hat{c}^\pdagger_{j\sigma} 
+  U \sum_{i} \hat{n}_{i\uparrow} \hat{n}_{i\downarrow}
\]
(with hopping amplitude $t$, onsite interaction $U$, and $\hat{n}_{i\sigma}=\hat{c}^{\dag}_{i\sigma}\hat{c}^\pdagger_{i\sigma}$)
can be realized to a reasonable accuracy using ultracold fermions in the interesting interaction range, which certainly supports the hopes of accessing also less understood Hubbard physics in similar ways.

However, all attempts of realizing and detecting {\it quantum magnetism} in cold lattice fermions have failed so far. In fact, it has not even been possible yet to verify specific signatures of antiferromagnetic (AF) correlations which are ubiquitous in correlated electrons and believed to play an important role in high-temperature superconductivity. This type of physics clearly has to be under control before cold fermions can really play a useful role as {\it quantum simulators}. Up to now the failures to detect AF signals have primarily been attributed to cooling issues \cite{Greif10,Santos10}. Indeed, the coldest systems achieved so far have central entropies per particle of $s \equiv S/(Nk_B) \approx \log(2)$ \cite{Joerdens10} while AF long-range order (LRO) on a cubic lattice is expected only for $s\lesssim \log(2)/2$ \cite{FWerner05,Wessel10,Greif10}.

In this Letter, we will argue that this discrepancy is not really relevant for the experiments currently performed or prepared in this context: both modulation spectroscopy \cite{Kollath06,Greif10} and the superlattice approach \cite{Trotzky10} address the nearest-neighbor (NN) spin correlation function $\NNcorr$ (for Pauli matrices $\hat{\boldsymbol{\sigma}}$). This is also true for the double occupancy $D\equiv\langle \hat{n}_{i\uparrow}\hat{n}_{i\downarrow}\rangle$ at large $U/t$ (and temperature $T=0$) \cite{Takahashi77}:
\[ 
D_{0}=\frac{Z t^2}{2U^2}\left(1-\NNcorr_0\right)+\mathcal{O}\left(\frac{t^4}{U^4}\right)
\]
However, as we will show using the example of $D$, all such observables are too local to be sensitive to LRO; given typical signal to noise ratios it seems very unlikely that the \neel\ transition could be detected in cold atoms in this way (assuming low enough $s$ is finally reached). As a consequence, full dimensionality (i.e. an isotropic cubic optical lattice) is not essential; in fact, a restriction of the atoms to planes or chains might enhance AF signals.

In the following, we will first briefly recall the DMFT scenario put forward in \cite{Gorelik_PRL10} and discuss arguments \cite{Fuchs11,De_Leo_PRA_2011} against the reliability of DMFT in dimension $d=3$. We will show, by comparisons with direct determinantal quantum Monte Carlo (QMC) simulations \cite{Blankenbecler81}, that the AF signatures predicted by DMFT survive even on the square lattice ($d=2$) and are surprisingly precise, up to rounding effects, in the cubic case ($d=3$). Finally, we will focus on a coupling strength corresponding to the ground state Mott transition \cite{Bluemer05} ($U/t=15$ for the cubic lattice) and demonstrate the effects of varying dimensionality from DMFT (exact for $d\to\infty$ \cite{fn:DMFT}) via $d=3$ and $d=2$ (from QMC) to $d=1$ (from a thermodynamic Bethe ansatz, BA \cite{BA}). 

%%%%%%%%%%%%%%%%%%%%%%%%%%%%%%%%%%%%%%%%%%%%%%%%%%%%%%%%%%%%%%%%%%%%%%%%%%%%%%%%%
% AF signatures in the double occupancy
%%%%%%%%%%%%%%%%%%%%%%%%%%%%%%%%%%%%%%%%%%%%%%%%%%%%%%%%%%%%%%%%%%%%%%%%%%%%%%%%%

\myparagraph{AF signatures in the double occupancy} 
A recent real-space DMFT study \cite{Gorelik_PRL10} showed that the low-temperature formation of an AF core in a fermionic cloud on an optical lattice (with half filling $n=1$ in the center) is signaled, at strong coupling, by a significant enhancement of $D$ in the same region. This DMFT scenario \cite{Gorelik_PRL10} is reproduced for $n=1$ in \reff{cubic_square}a: 
%%%%%%%%%%%%%%%%%%%%%%%%%%%%%%%%%%%%%%%%%%%%%%%%%%%%%%%%%%%%%%%%%%%%%%%%%%%%%%%%%
\begin{figure}[t]
\includegraphics[width=\columnwidth]{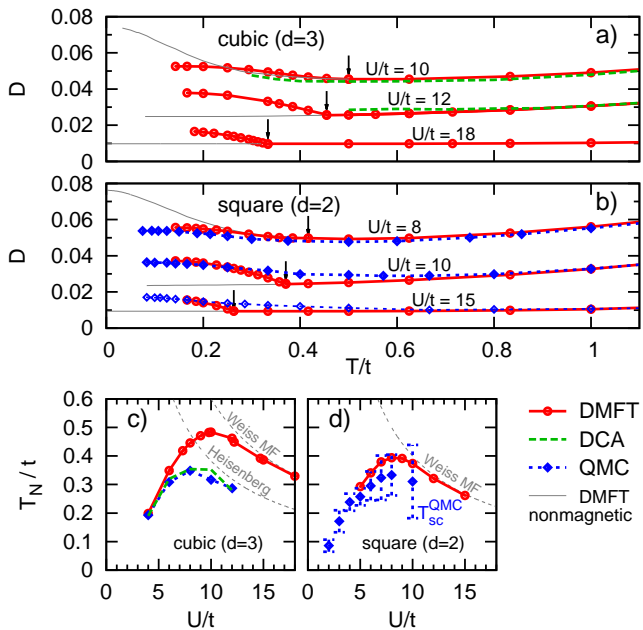}
\caption{(Color online)
Top: Double occupancy $D(T)$ as estimated from DMFT (circles), QMC (diamonds), and DCA \cite{Fuchs11} (dashed lines) for (a) cubic and (b) square lattice. Thin lines: nonmagnetic DMFT results for $T<\TND$ (arrows).\\
Bottom: (c) The \neel\ temperature $T_N(U)$ (determined by QMC \cite{Staudt00} and DCA \cite{Kent05}) is up to $30\%$ below the DMFT estimate for $d=3$. (d) While $T_N$ vanishes in $d=2$, the spin-crossover temperature \cite{Paiva10} agrees with $T_N^{\text{DMFT}}(U)$.
}\label{fig:cubic_square}
\end{figure}
%%%%%%%%%%%%%%%%%%%%%%%%%%%%%%%%%%%%%%%%%%%%%%%%%%%%%%%%%%%%%%%%%%%%%%%%%%%%%%%%%
at strong coupling $U/t=18$ the double occupancy $D(T)$ is asymptotically flat at low $T$ in the nonmagnetic phase, but is strongly enhanced, by up to $75\%$, when AF order sets in below $\TND\approx 0.35t$ (arrow). The relative enhancement quickly decreases at smaller $U$ and is lost at $U/t\approx 10$. The absolute enhancement of $D$ is largest for $U/t\approx 12$ and should be detectable experimentally even in measurements integrating over the whole cloud \cite{Gorelik_PRL10}.

This scenario was challenged recently \cite{Fuchs11} on the basis of the dynamical cluster approximation (DCA) which relaxes the DMFT assumption of a momentum independent self-energy \cite{Maier_RMP_2005}: the DCA estimates of $D$ (dashed lines in \reff{cubic_square}a) showed no clear AF related enhancement \cite{Fuchs11}; however, these calculations could not enter the low-$T$ AF phase. The reliability of DMFT estimates for $D$ and $s$ (in the nonmagnetic phase) at low $T$ was also questioned based on comparisons with high-temperature expansions (HTE) \cite{De_Leo_PRA_2011}. It is indeed clear that the DMFT scenario cannot be correct in all aspects: after all, it is well-known that DMFT overestimates the \neel\ temperature by up to $30\%$ in 3 dimensions (see \reff{cubic_square}c). Thus, the kinks in $D(T)$ at $\TND$ are certainly unphysical -- but is the whole scenario just a DMFT artifact?

%%%%%%%%%%%%%%%%%%%%%%%%%%%%%%%%%%%%%%%%%%%%%%%%%%%%%%%%%%%%%%%%%%%%%%%%%%%%%%%%%
% Comparison in $d=2$
%%%%%%%%%%%%%%%%%%%%%%%%%%%%%%%%%%%%%%%%%%%%%%%%%%%%%%%%%%%%%%%%%%%%%%%%%%%%%%%%%

\myparagraph{Comparison in $d=2$} 
For a first answer, let us turn to the square lattice ($d=2$) for which the DMFT is {\em a priori} much less reliable than in $d=3$. In fact, DMFT predicts AF LRO even in this case, with a maximum in $\TND$ of about $0.4t$ at $U/t\approx 8$ (circles in \reff{cubic_square}d), while the Mermin-Wagner theorem excludes LRO for $T>T_N=0$. However, in this case it is relatively easy to check the DMFT predictions (circles in \reff{cubic_square}b) by direct QMC simulations \cite{Blankenbecler81} of finite clusters, here of size $10\times 10$ (diamonds). After employing an (approximate) correction for Trotter errors and verifying that finite-size effects are negligible we consider this data essentially exact. The previously established accuracy of the DMFT at high temperatures \cite{De_Leo_PRA_2011,Fuchs11} evidently survives in $d=2$, with no significant deviations from QMC for $T/t\gtrsim 0.8$. Surprisingly good agreement is also found at low temperatures $T/t\lesssim 0.2$, although the stable DMFT solutions (circles) here correspond to the AF phase which at first sight appears unphysical. In contrast, DMFT calculations constrained to the nonmagnetic phase (thin lines) predict low-$T$ features of $D$ which are far off from the exact QMC data. This teaches an important lesson, relevant also for $d=3$: paramagnetic phases which include short range order can be much more similar to AF phases (with AF LRO) than to nonmagnetic solutions (without any AF correlations). Of course, DMFT is still not perfect: QMC shows significant corrections of DMFT predictions, namely a rounding of the unphysical kinks, at $T\approx \TND$. However, even $\TND$ has physical significance: it (nearly) matches the spin coherence temperature (diamonds in \reff{cubic_square}d) \cite{Paiva10}. 

%%%%%%%%%%%%%%%%%%%%%%%%%%%%%%%%%%%%%%%%%%%%%%%%%%%%%%%%%%%%%%%%%%%%%%%%%%%%%%%%%
% Comparison in $d=3$
%%%%%%%%%%%%%%%%%%%%%%%%%%%%%%%%%%%%%%%%%%%%%%%%%%%%%%%%%%%%%%%%%%%%%%%%%%%%%%%%%

\myparagraph{Comparison in $d=3$} 
QMC results on a cubic lattice, the target system of current AF related experiments, were obtained for 
clusters with $6^3$ and $8^3$ sites and carefully extrapolated to vanishing Trotter discretization $\dt\to 0$ (large diamonds in \reff{QMC_comparison}).
%%%%%%%%%%%%%%%%%%%%%%%%%%%%%%%%%%%%%%%%%%%%%%%%%%%%%%%%%%%%%%%%%%%%%%%%%%%%%%%%%
\begin{figure}[t]
\includegraphics[width=\columnwidth]{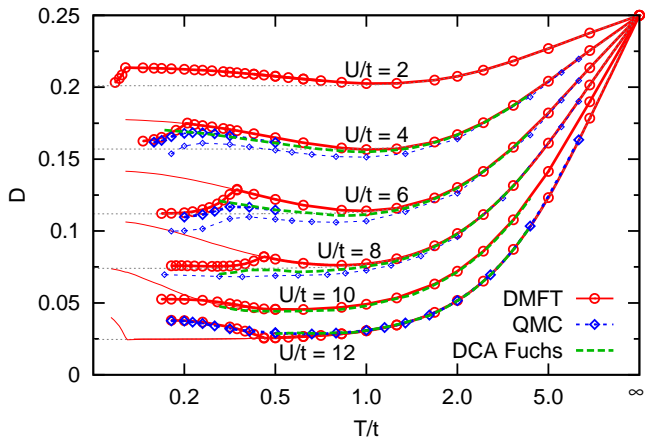}
\caption{(Color online)
Cubic lattice ($d=3$): 
Comparison of DMFT (circles) and direct QMC (diamonds) results for the temperature dependence of the double occupancy $D(T)$ for the cubic lattice. DCA results taken from \cite{Fuchs11}. Small diamonds: QMC results with finite Trotter discretization bias. 
}\label{fig:QMC_comparison}
\end{figure}
%%%%%%%%%%%%%%%%%%%%%%%%%%%%%%%%%%%%%%%%%%%%%%%%%%%%%%%%%%%%%%%%%%%%%%%%%%%%%%%%%
These data show perfect agreement with the DMFT estimates (circles) at $U/t=12$ both for $T/t\ge 0.7$ and for $T/t\le 0.4$; thus the DMFT prediction of the $D$ enhancement \cite{Gorelik_PRL10} is even quantitatively correct. Only at $\TND\approx0.45t$ the QMC results smooth out the DMFT kink.

At relatively weak coupling $U/t\le 8$, signatures appear in the DMFT data in \reff{QMC_comparison}  which differ fundamentally from the strong-coupling scenario discussed so far: $D(T)$ shows a broad minimum at $T\approx t$; the rise towards lower $T$ breaks down quite abruptly below $\TND$, remarkably approaching exponential fits to the high-$T$ behavior (dotted lines). Apparently the system behaves as a (bad) insulator for $T\gtrsim t$; the Fermi liquid behavior setting in for $T\lesssim t$ enhances $D$ \cite{FWerner05}, but is destroyed below $\TND$ by AF correlations. Also this weak-coupling DMFT scenario for $D(T)$ is confirmed: QMC predicts (large diamonds) a peak right at $\TND\approx 0.2t$ for $U/t=4$  and quickly converges towards DMFT for lower $T$. The deviations in the range $\TND\lesssim T\lesssim t$ can be traced to developing AF correlations which already reduce the Fermi liquid enhancement of $D$. Note that (at $U/t=4$) the discrepancies between QMC and DMFT are much smaller than typical QMC discretization errors (data for $\dt\, t=1/8$: small diamonds) and that DCA (dashed line) apparently misses the AF physics at $T/t\lesssim 0.2$. 

LRO leaves traces in the QMC estimates of $D(T)$ only in the weak-coupling regime $U/t\lesssim 6$ where $\TN\approx \TND$ (cf.\ \reff{cubic_square}c). At strong coupling $U/t=12$, $D(T)$ does not show visible features at $\TN\approx 0.3t$, which suggests that local spin correlations (which determine $D$ and current AF observables \cite{Greif10,Trotzky10}) are hardly sensitive to LRO and, consequently, dimensionality in this regime.

%%%%%%%%%%%%%%%%%%%%%%%%%%%%%%%%%%%%%%%%%%%%%%%%%%%%%%%%%%%%%%%%%%%%%%%%%%%%%%%%%
% Impact of dimensionality and entropy
%%%%%%%%%%%%%%%%%%%%%%%%%%%%%%%%%%%%%%%%%%%%%%%%%%%%%%%%%%%%%%%%%%%%%%%%%%%%%%%%%

\myparagraph{Impact of dimensionality and entropy}
In order to gain more insight into these issues, DMFT and QMC data for the cubic lattice are compared at $U/t=15$ with QMC results for the square lattice and BA solutions of the infinite chain at (nearly) equivalent \cite{fn:scale} interactions in \reff{3d_SvsT}a.
%%%%%%%%%%%%%%%%%%%%%%%%%%%%%%%%%%%%%%%%%%%%%%%%%%%%%%%%%%%%%%%%%%%%%%%%%%%%%%%%%
\begin{figure}[t]
\includegraphics[width=\columnwidth]{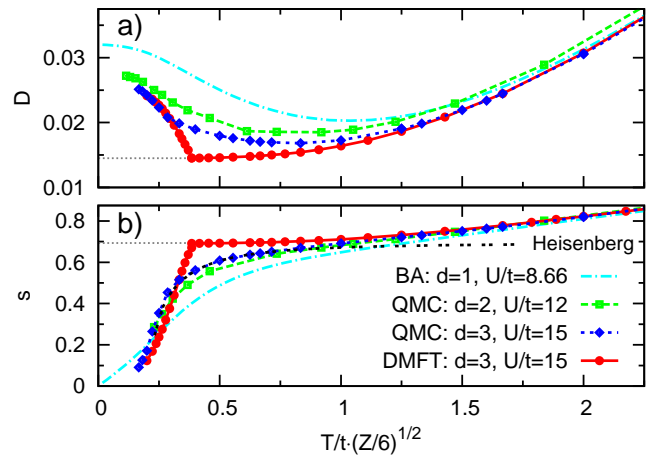}
\caption{(Color online)
Hypercubic lattice ($1\le d\le 3$) at strong coupling:
a) $D(T)$ as estimated from  DMFT ($d=3$, circles), QMC ($d=2$ \cite{fn:2d}, $3$, diamonds), and BA ($d=1$, dash-dotted line). 
b) Corresponding estimates of  entropy per particle $s=S/N$. All interactions correspond approximately to the ground state Mott transition at $U/(\sqrt{Z}t)\approx 6$.
}\label{fig:3d_SvsT}
\end{figure}
%%%%%%%%%%%%%%%%%%%%%%%%%%%%%%%%%%%%%%%%%%%%%%%%%%%%%%%%%%%%%%%%%%%%%%%%%%%%%%%%%
Here, the DMFT data (circles) can also be interpreted as an exact result in infinite dimensions. After rescaling \cite{fn:scale}, we find rapid convergence with increasing dimensionality at high $T$ and generally similar shapes \cite{fn:2d} of $D(T)$ for $1\le d\le \infty$. However, $d\gg 3$ would apparently be needed in order to converge to the DMFT results also at $\TND$. Furthermore, the minimum in $D(T)$ occurs at about twice $\TND$ in dimensions $1\le d\le 3$, reinforcing doubts about the usefulness of DMFT estimates of $D$ for thermometry \cite{Fuchs11}.

It is well-known that nonmagnetic DMFT yields an entropy $s \xrightarrow{T\to 0} \log(2)$ (dotted line in \reff{3d_SvsT}b), which is clearly unphysical \cite{De_Leo_PRA_2011}. However, the AF DMFT solution (circles for $T<\TND$) recovers the QMC results for the cubic lattice (diamonds) at $T\lesssim \TN \approx 0.25t$; remarkably the latter coincide with the Heisenberg limit of $s(T)$ for $T\lesssim 0.8t$. In general, the dimensional dependence of $s(T)$ nearly mirrors that of $D(T)$. Thus, dimensional effects and DMFT errors should be minimal when using $s$ as a  (dimensionless) measure of temperature (which is of primary interest to experimentalists anyway).

Indeed, as seen in \reff{dim_DvsS}, $D(s)$ looks strikingly similar in all dimensions; in particular, 
%%%%%%%%%%%%%%%%%%%%%%%%%%%%%%%%%%%%%%%%%%%%%%%%%%%%%%%%%%%%%%%%%%%%%%%%%%%%%%%%%
\begin{figure}[t]
\includegraphics[width=\columnwidth]{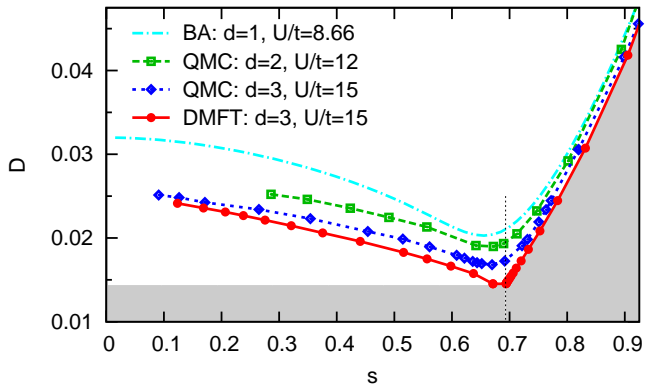}
\caption{(Color online)
Hypercubic lattice ($1\le d\le 3$) at strong coupling: 
Double occupancy as a function of entropy per particle. In all cases, the minimum of the double occupancy corresponds to $s\approx \log(2)$ (dotted line). The shaded area indicates the nonmagnetic contribution to $D$.
}\label{fig:dim_DvsS}
\end{figure}
%%%%%%%%%%%%%%%%%%%%%%%%%%%%%%%%%%%%%%%%%%%%%%%%%%%%%%%%%%%%%%%%%%%%%%%%%%%%%%%%%
the minimum in D (at strong coupling) corresponds to $s\approx \log(2)=s_{\text{N}}^{\text{DMFT}}$ in all cases! While it might appear surprising that this behavior persists down to $d=1$ it is clear that $s< \log(2)$ is only possible for a two-flavor system at $n=1$ by spin coherence, i.e. the development of (possibly short ranged) AF correlations; these, in turn, enhance $D$ \cite{Gorelik_PRL10}. 

Thus, the evolution of $D$ is a near-perfect thermometer for ultracold atoms measuring AF correlations and (in $d=3$) the proximity to AF LRO. In fact, we would argue that any positive deviation of $D(s)$ from the nonmagnetic background (shaded in \reff{dim_DvsS}) originates from AF correlations which are strong for $s\lesssim \log(2)\approx 0.7$ in all dimensions (and coincide with LRO in $d=\infty$ and, at $s<s_N\approx \log(2)/2$, in $d=3$) and are strictly zero only in the limit $d\to\infty$ for $s\ge \log(2)$.  This enhancement is larger in lower $d$, consistent with the known dimensional dependence of $\NNcorr$ for the Heisenberg model at $T=0$: $\NNcorr_0 = -1.00$ ($d=\infty$, Weiss MF); $-1.20$ ($d=3$) \cite{Oitmaa94}; $-1.34$ ($d=2$) \cite{Sandvik97}; $-1.77$ ($d=1$) \cite{Takahashi77}. Thus, irrespective of the measurement technique, signatures of AF correlations may be easier to detect experimentally (at fixed $s$) for lower (effective) dimensionality. Conversely, a tuning of the hopping amplitude in $z$ direction could help to discriminate magnetic effects from those of charge excitations; similar ideas including frustration will be explored in a separate publication \cite{Frustration}.

%%%%%%%%%%%%%%%%%%%%%%%%%%%%%%%%%%%%%%%%%%%%%%%%%%%%%%%%%%%%%%%%%%%%%%%%%%%%%%%%%
% Conclusion
%%%%%%%%%%%%%%%%%%%%%%%%%%%%%%%%%%%%%%%%%%%%%%%%%%%%%%%%%%%%%%%%%%%%%%%%%%%%%%%%%

\myparagraph{Conclusion} 
In this Letter, we have demonstrated that DMFT predicts temperature dependencies of local quantities, e.g. $D(T)$, more accurately in $d=3$, $d=2$ than expected. Especially, both the AF induced enhancement (at strong coupling), and the suppression (at weak coupling) of $D$ survive. The temperature scale given by $\TND (U)$ corresponds to a spin-crossover in finite dimensions. As a function of entropy per particle $s$, the double occupancy is nearly universal with respect to dimensionality; in particular, the minimum in $D(s)$ always occurs at $s\approx \log(2)$ at strong coupling, as predicted by DMFT. 
Thus, we have established a prominent and specific signal of AF correlations in an entropy range that is in immediate experimental reach, with the prospect of extending the use of $D(s)$ for thermometry \cite{Joerdens10} to the most interesting range $s\lesssim log(2)$. Our results also validate the RDMFT approach \cite{Bluemer11CCP} for quantitative simulations of inhomogeneous 3-dimensional systems. 

We thank M.\ Inoue for help with the BA code, and P.G.J.\ van Dongen, U.\ Schneider, and R.\ P.\ Singh for valuable discussions. Support under ARO Award W911NF0710576 with funds from the DARPA OLE Program, by CNPq and FAPERJ, and by the DFG within SFB/TRR 49 is  gratefully acknowledged.

%%%%%%%%%%%%%%%%%%%%%%%%%%%%%%%%%%%%%%%%%%%%%%%%%%%%%%%%%%%%%%%%%%%%%%%%%%%%%%%%%
% Bibliography
%%%%%%%%%%%%%%%%%%%%%%%%%%%%%%%%%%%%%%%%%%%%%%%%%%%%%%%%%%%%%%%%%%%%%%%%%%%%%%%%%


\begin{thebibliography}{16}
\bibitem{fn:DMFT} 
The DMFT impurity problem is solved using the Hirsch-Fye QMC algorithm with (nearly) exact elimination of the Trotter errors by extrapolation $\dt\to 0$ \cite{Bluemer07}.
\bibitem{fn:scale}
All scales are set by the root mean square energy $\sqrt{\langle \epsilon^2\rangle_{U=0}}=\sqrt{Z} t$ (for coordination number $Z$), not by the band edges ($\sim\! Z$): the DMFT data of Figs.\ \ref{fig:cubic_square}a and \ref{fig:cubic_square}b nearly matches after rescaling $U$ and $T$ by $\sqrt{6/4}\approx 1.22$.
\bibitem{fn:2d}
Deviations for $d=2$ in Figs.\ \ref{fig:3d_SvsT} and \ref{fig:dim_DvsS} are due to a slightly too low interaction 
%$U/t=12$ instead of the scaling value $15/\sqrt{6/4}\approx 12.25$.
$U/t=12<15/\sqrt{6/4}\approx 12.25$.
\end{thebibliography}
\end{document}